\documentclass{aa}
\usepackage{graphics}

\def\kms    {\ifmmode{{\rm ~km~s}^{-1}}\else{~km~s$^{-1}$}\fi}

\def\arcm   {$^{\prime}$}
\def\arcmper  {\ifmmode \rlap.{' }\else $\rlap{.}' $\fi}

\def\arcsper  {\ifmmode \rlap.{'' }\else $\rlap{.}'' $\fi}
\def\arcsgper  {\ifmmode \rlap.^{s }\else $\rlap{.}^s $\fi}   
\def\ddeg     {{\rlap.}$^{\circ}$}
\def\deg      {\ifmmode^\circ\else$^\circ$\fi}     

\def\solmass  {M$_\odot$}
\def\sper     {\ifmmode \rlap.^{s}\else $\rlap{.}^s$\fi}

\def\>           {$>$}
\def\<           {$<$}

\def\kms    {\ifmmode{{\rm ~km~s}^{-1}}\else{~km~s$^{-1}$}\fi}

\begin{document}
\thesaurus{11     
              (11.05.2;  
               11.09.2
               11.11.1;  
               11.19.6;  
               13.19.1;  )}

\title{Where is the Neutral Atomic Gas in Hickson Groups?\footnote{This work is
partially based on
observations made with the VLA operated by the National Radio
Astronomy Observatory, a
facility of the National Science Foundation operated under cooperative
agreement by Associated Universities, Inc., 
ALFOSC, which is owned
by the Instituto de Astrof\'{\i}sica de Andaluc\'{\i}a (IAA, CSIC) and
operated at the Nordic Optical Telescope (NOT) under agreement between IAA
and the NBIfA of the Astronomical Observatory of Copenhagen, and 1.5m
telescope of the Observatorio de Sierra Nevada, Granada, Spain, which is
operated by the IAA (CSIC).}}
 
\author{
L. Verdes-Montenegro \inst{1}
\and
M. S. Yun \inst{2}
\and
B. A. Williams \inst{3}
\and
W. K. Huchtmeier  \inst{4}
\and
A. Del Olmo  \inst{5}
\and
J. Perea \inst{6}
}
   \offprints{L. Verdes-Montenegro           }
\institute{
Instituto de Astrof\'{\i}sica de
Andaluc\'{\i}a, CSIC, Apdo. Correos 3004, E-18080 Granada, Spain,
lourdes@iaa.es
\and
Astronomy Department. University of Massachusetts. 
Amherst, MA 01003,USA
 myun@astro.umass.edu
\and
University of Delaware, Newark, Delaware, USA
baw@udel.edu
\and
Max-Planck-Institut f\"{u}r
Radioastronomie
Auf dem H\"{u}gel 69, D-53121 Bonn, Germany
huchtmeier@mpifr-bonn.mpg.de
\and
Instituto de Astrof\'{\i}sica de
Andaluc\'{\i}a, CSIC
Apdo. Correos 3004, E-18080 Granada, Spain
chony@iaa.es
\and
Instituto de Astrof\'{\i}sica de
Andaluc\'{\i}a, CSIC
Apdo. Correos 3004, E-18080 Granada, Spain
jaime@iaa.es
}
\date{ }
\titlerunning{Where is the HI in HCGs?}
\authorrunning{Verdes-Montenegro et al.}    

\maketitle

\begin{abstract}

We have analyzed the total HI contents of 72 Hickson 
compact groups of galaxies (HCGs) and the detailed spatial 
distributions and kinematics of HI within a subset of 16 groups
using the high angular resolution observations obtained with
the VLA in order to investigate a possible evolutionary 
scenario for these densest systems in the present day 
galaxy hierarchy.  For the more homogeneous subsample of 48 groups,   
we found a mean HI deficiency of 
{\it Def$_{\rm HI}$} = 0.40 $\pm$ 0.07, which corresponds
to 40\% of the expected HI for the optical luminosities and 
morphological types of the member galaxies.
The individual galaxies show larger degrees of deficiency than the  
groups globally, {\it Def$_{\rm HI}$} = 0.62 $\pm$ 0.09 (24\% of the expected HI), 
due in most cases to
efficient  gas stripping  from individual galaxies into the group environment
visible in the VLA maps.
The degree of deficiency is found
to be similar to the central galaxies of Virgo and Coma cluster, and Coma 
I group, in spite of the significantly different characteristics 
(number of galaxies, velocity dispersion) of these environments. 
It does not seem plausible that a significant
amount of extended HI has been missed by the observations.
Hence phase transformation of the atomic gas 
should explain the HI deficiency. 
The groups richer in  early type galaxies or more compact with 
larger velocity dispersions show a weak tendency to be more HI deficient.
The detection rate of HCGs at X-ray wavelengths is larger for 
HI deficient groups, although the hot gas distribution and hence 
its origin is only known for a few cases. 
In the evolutionary scenario we propose, the amount of detected HI
would decrease further with evolution, by continuous tidal
stripping and/or heating.

The H$_{\rm 2}$ content also tends to be lower than expected for the 
galaxies in HI deficient groups, this may suggest that
the HI stripping by frequent tidal interaction 
breaks the balance between the disruption of
molecular clouds by star formation and the replenishment from the 
ambient HI.

  \end{abstract}

\section{Introduction}

The Hickson Compact Groups (Hickson 1982: HCGs) are characterized by  
a small number of members (4 to 10) with a low
velocity dispersion ($<$ $\sigma$ $>$ = 200 km~s$^{-1}$, Hickson et al
1992).
Their projected galaxy densities are extremely high,
similar to the cores of dense clusters, but they are found
in low galaxy density environments, as a consequence of the
isolation criterion used for their selection (Hickson 1982; Sulentic 1987).
This combination of high galaxy density in the low density environment
makes them unique and interesting laboratories for studying 
galaxy interaction and evolution.

The origin and the existence of Hickson compact groups constitute 
a matter of great interest and debate.  Their short
crossing times argue for short lifetimes, yet few 
merger candidates or postcursors 
are found (Zepf et al. 1991; Moles et al. 1994; Sulentic \& Rabaça 1994).
Diaferio et al. (1994) proposed that 
compact groups could form continually out of rich groups.
Governato et al. (1996) suggest that primordial merger events 
with ongoing acquisition of intruders might 
prevent the groups from merging entirely.
These models seem at work in HCG 92 (Moles et al. 1997).
On the other hand, well isolated groups with no obvious intruders 
also exist (e.g. HCG 96 Verdes-Montenegro et al. 1997).
Athanassoula et al. (1997) have shown that compact groups with 
an appropriate arrangement of luminous and dark matter can 
persist over several hundred million years, in agreement with the 
analysis of the dynamics of satellite galaxies (Perea et al. 2000).

The characteristics of the interstellar medium in HCGs are also
not well
understood.  Elevated star formation activity induced
by tidal interactions and resulting enhancement in FIR and CO 
emission are expected but not supported by observations 
(see  Verdes-Montenegro et al. 1998 and references therein). 
The highly perturbed distributions of the molecular gas in the most
CO deficient groups suggest that strong disruptions of gaseous
disks and gas stripping by continuous tidal disruption may 
suppress star formation (Yun et al. 1997). 
 
A detailed analysis of the morphology and the strength of interactions 
among the HCG galaxies is needed in order to better understand the 
way these compact groups form and evolve.  Studies of neutral 
hydrogen (HI) can offer valuable insights into the dynamics
of galaxy interaction and evolution (e.g. Yun et al. 1994,
Hibbard \& van Gorkom 1997), and we were motivated to undertake 
a survey of HI distribution and content for a large sample of HCGs
(Williams et al. 1997, 1999; Verdes-Montenegro et al. 2000ab; 
Huchtmeier et al. 2000) in order to analyze the type and effect of 
the interactions that are taking place.
In this paper we report the analysis of the HI content for 72 HCGs 
and combine this with the analysis of the high spatial and spectral
resolution VLA HI observations for a subset of 16 groups. 
Detailed studies of the individual groups will be presented
elsewhere (in preparation).  
We have determined the frequency of perturbed groups from the
HI content, the morphology, and the nature of the anomalies 
in the HI distribution.
A global HI deficiency in HCGs has been  previously reported
in the literature (Williams \& Rood 1987; Huchtmeier 1997). 
We explore here a large set of parameters that might account for 
the HI anomaly. The high
spatial and kinematical resolution of the HI data should help 
restrict the boundary conditions used in hydrodynamical modeling of galaxy
interactions.

\section{The Sample Selection and the HI Data \label{sec:sample}}

\subsection{Single-Dish Sample \label{sec:SDsample}}

We have investigated  published data for all groups from the  
Hickson catalog (Hickson
1982) containing at least 3 concordant
members (hence HCG 9, 11, 36, 41 and 77 have been excluded; Hickson 
et al 1992). 
HCG 70 is a projected superposition of a triplet and a quadruplet at 
different redshifts.  Hence there are a total of 96 true groups in
the Hickson Catalog.   The HI single-dish spectra are available
for 72 groups, either
from the Arecibo and Green Bank radiotelescopes (Williams \& Rood 1987)
and/or from the 100m Effelsberg antenna (Huchtmeier 1997), and
a total of 58 groups are detected.   There is good agreement in the 
fluxes of the 31 groups with duplicate measures,  
as indicated in Huchtmeier (1997). Williams \& Rood (1987) find 
that, on average, HCGs are 50\% deficient in HI compared
to a  sample of loose groups.
Huchtmeier (1997) finds a large scattering in the ratios of integrated
HI mass to optical luminosity, 
including strong HI deficiencies for a large number of objects.
 
\subsection{The VLA Sample \label{sec:VLAsample}}

We have mapped a total of 16 HCGs using the VLA.  Detailed analysis
of six spiral dominated groups have already been published
(HCG 18, Williams 
\& Van Gorkom 1988; HCG 31, 44 and 79,  Williams et al. 1991; HCG 23 and 26,
Williams \& Van Gorkom 1995), and the preliminary analysis for the 
5 other spiral dominated groups have also been reported 
(Williams et al. 1997, 1999; Verdes-Montenegro et al. 2000ab).
HCG 92 (Stephan's Quintet) was also previously studied at WSRT 
(Shostak et al. 1984).
We recently obtained the VLA data for five new groups (HCG 40, 49,
54, 95 and 96) in an effort to expand the sample to include
more representatives of different interaction scenarios found in HCGs.
The new additions include the densest compact groups, where the 
strongest interactions are expected, as well as some
groups that are known to be HI deficient, as HCG 40, 44 and 92. 
A summary of the parameters for all observations is given in 
Table~\ref{tab:vlasummary}.
Single-dish fluxes are compared to the integrated
fluxes for the 16 groups observed with the VLA (corrected for primary beam
attenuation, and removing the contributions from neighbor galaxies
detected in HI within the VLA beam but external to the single-dish beam).
Only in
two cases (HCG 33 ad HCG 95) 
are we concerned that the interferometer
may have missed flux (about 50\%).
On the other hand the VLA flux for HCG 16 is larger than the measured
value with a single dish, due to an HI extent larger than the single-dish beam.

\begin{table*}
\caption{Observational parameters of the interferometric observations.}
\label{tab:vlasummary}
\begin{center}
\begin{tabular}{lcccccc}
HCG& VLA& \multicolumn{2}{c}{Beam size}& Channel& 
$\sigma(S_\nu)$ & $\sigma(N_{\rm HI})^{a}$ \\ 
&  configuration& \multicolumn{2}{c}{}& width & & \\ 
& & ($'' \times ''$) & (kpc $\times$ kpc)& (km/s) & (mJy/beam)& 
($10^{18}$ cm$^{-2}$) \\ 
\hline 
& & & & & &\\ 
2 & D& 68 $\times$ 58& 18.0 $\times$ 16.2& 10.6& 0.64& ~1.9\\ 
16& C+D& 25 $\times$ 19 & 6.3 $\times$ 5.0& 21.3 & 0.31& 15.0\\ 
18&C& 20 $\times$ 20& 5.2 $\times$ 5.2&21.2&0.50& 28.6\\ 
23&C& 21 $\times$ 17& 5.3 $\times$ 6.6&10.6&1.70&54.5 \\ 
26&C& 25 $\times$ 18& 15.3 $\times$ 11.0&21.8&0.40& 20.9\\ 
31&CnD&21 $\times$ 18& 5.6 $\times$ 4.8&21.0&0.22& 13.2\\ 
&CnB& 16 $\times$ 14& 4.2 $\times$ 3.9&10.6&1.50& 76.6\\ 
33& C &18 $\times$ 16& 9.0 $\times$ 8.3& 21.8 & 0.65 & 53.1\\ 
40& C& 23 $\times$ 18& 10.1 $\times$ 7.9& 10.8&0.65& 18.3\\ 
44&D& 61 $\times$ 60& 5.4 $\times$ 5.4&42.0&0.40& ~5.0\\
49& C& 41 $\times$ 23& 26.3 $\times$ 14.8& 11.0&0.51& ~6.4\\
54& C& 20 $\times$ 16& 1.9 $\times$ 1.5& 10.4&0.33& 11.6\\ 
79 & CnD& 22 $\times$ 17& 7.1 $\times$ 4.5&21.0&0.86& 52.1\\ 
&C& 25 $\times$ 16& 7.0 $\times$ 4.5&10.6&0.39& 11.2\\
88& C& 24 $\times$ 17& 9.2 $\times$ 6.8&21.4&0.43& 24.3\\ 
92& C+Cs+D& 20 $\times$ 19& 7.8 $\times$ 7.3&21.5&0.54& 33.0\\ 
95& C+D&23 $\times$ 16& 17.7 $\times$ 12.4&22.2&0.22& 14.3\\ 
96&C+D& 33 $\times$ 25& 18.7 $\times$ 14.3&21.9&0.34& ~9.7\\ 
\\
\hline
\end{tabular} 
\begin{list}{}{} 
\item[$^{\rm a}$] 
Column density obtained integrating for a channelwidth.
\end{list} 
\end{center}
\end{table*}

HCG 22, 26, 42, 48, 63, 65 and 91 have been 
observed with  the ATNF interferometer, but maps are only published for
HCG 22 and 26 (Price et al. 2000). The sensitivity of the ATNF observations
of HCG 26 is
considerably lower than for the VLA data. 
No information
on the total recovered flux is given for HCG 22, so 
it is not possible to evaluate how representative
the VLA flux distribution is of the total HI emission.

\section{HI content of the HCGs \label{sec:HImass}}

\subsection{The single-dish sample \label{sec:MHISD}}
 
The atomic hydrogen mass  of the 72 groups with single-dish data 
is computed as $M(HI) = 2.36 \times 10^5~D^2 S \Delta V$,
where $D$ is the luminosity distance in Mpc (assuming
$H_{\rm 0} = 75 km  s^{-1}Mpc^{-1}$) and $S \Delta V$ is the velocity
integrated HI flux in Jy {\rm ~km~s}$^{-1}$.  Fluxes 
are taken from 
Williams \& Rood (1987) and Huchtmeier (1997).
For each galaxy the predicted mass is calculated as a function of its
optical luminosity and morphological type (taken from Hickson et al
1989) via the relationships obtained by Haynes \& Giovanelli (1984)
for a sample of 324 isolated galaxies.  
In fact 34 galaxies were excluded from their analysis since they 
were not detected. Those correspond mostly to early types, and for this
reason larger errors can be expected in the relationships. This
effect is considered in Sect. 4.
The HI mass expected for each group is obtained as the sum of the 
calculated masses for all its spiral and lenticular members. 

The HI deficiency is defined
as {\it Def$_{\rm HI}$} = log[M(HI)$_{\rm pred}$]--log[M(HI)$_{\rm obs}$].
We also adopt the definition of the standard estimate of the error (s.e.e.) 
in the HI masses predicted from the optical luminosity 
L$_{\rm B}$ as defined by Haynes \& Giovanelli (1984). 
This is on the order of log[M(HI)] = 0.20 for the HCGs studied.
We consider the HI content of a galaxy/group anomalous when 
its HI content deviates
from the predicted value by more than twice the mean s.e.e. 
The expected HI masses for the groups were 
derived as 
sums of the expected HI masses for
the individual member galaxies. 
This approach is used because it takes into account
the morphological dependency of M(HI) and 
the well known non-linearity in the M(HI) --
L$_{\rm B}$ relation (M(HI) $\sim$ L$_{\rm B}^{0.63 - 0.84}$; Haynes \&
Giovanelli 1984). 
Finally it would imply in many cases 
to extrapolate the M(HI) versus L$_{\rm B}$ relationships above the range of
values used in their determination.

The observed and predicted HI masses are compared in Fig. 1  (the 
diagonal line indicates  the locus of galaxies with 
normal HI content) and 
are listed in Table~\ref{tab:groupmass} along with the s.e.e.
An histogram of the HI deficiency for the 72 groups, shown 
in the inset of Fig. 1, gives a mean value of 
{\it Def$_{\rm HI}$} = 0.36 $\pm$ 0.06.  The upper and lower limits have 
been taken into account by means of survival analysis. 
This value will be revised in  Sect. 4 considering
effects that might increase/decrease the HI deficiency.

\begin{figure*}
\resizebox{\hsize}{!}{\includegraphics{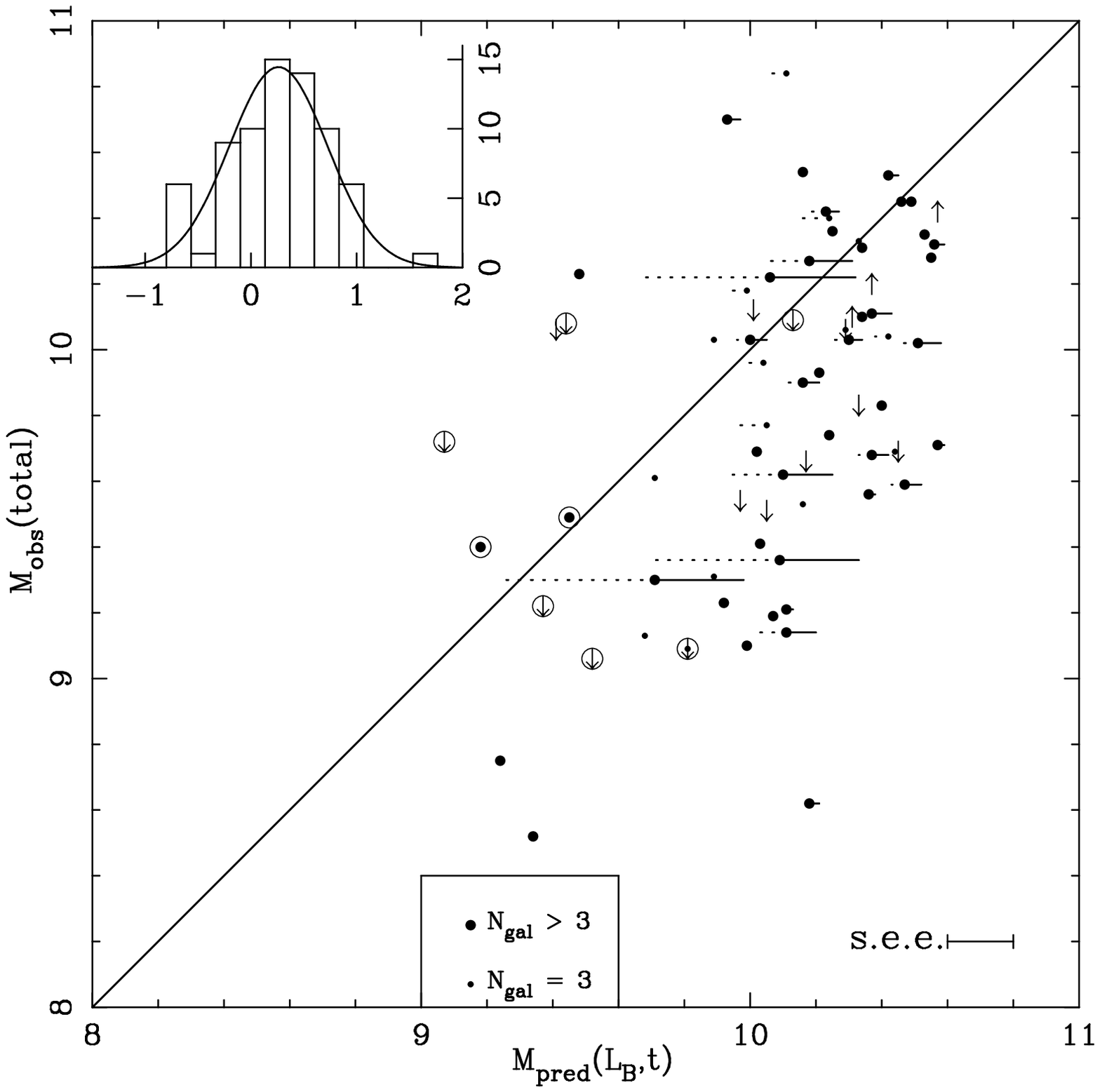}}
\hfill
\caption{Comparison of the total observed HI mass in each
HGC with HI single-dish data   with the expected one obtained 
as described in \S~\ref{sec:MHISD}.  
Triplets are marked with smaller symbols.
Dashed lines indicate the decrease in expected HI mass if one 
assumes  lenticulars 
to be devoid of HI. Solid lines  correspond to  the increase
in the predicted mass of groups containing lenticular galaxies
assuming that all were originally Sbs.   
Arrows represent upper and lower limits. 
Big open circles indicate groups with no spiral members.          
The segment in the lower right indicates the mean value of the 
standard estimate of the error (s.e.e.). 
The solid line represents a normal HI content.
An histogram of the HI deficiency for the 72 groups of the sample is shown 
in the inset in the upper left of the figure. }
\label{fig:MHISD}
\end{figure*}

\begin{table*}
\caption{HI masses for groups.}
\label{tab:groupmass}
\begin{tabular}{rrrrcl}
HCG& Distance &
\multicolumn{1}{c}{log[M(HI)$_{\rm obs}$]}&
\multicolumn{1}{c}{log[M(HI)$_{\rm pred}$]$^{a}$}&
s.e.e.&
Remarks\\
& (Mpc)&
\multicolumn{1}{c}{(log[M$_{\odot}$])}&
\multicolumn{1}{c}{(log[M$_{\odot}$])}&
(log[M$_{\odot}$])& \\  
  1&   136.6&    10.53~~~~~& 10.42~~~~~~~~~~&  0.19& \\     
  2&    57.8&    10.33~~~~~& 10.33~~~~~~~~~~&  0.19& triplet \\    
  3&   102.6&    10.18~~~~~& 9.99~~~~~~~~~~&  0.19& triplet \\    
  4&   112.7&    10.31~~~~~& 10.34~~~~~~~~~~&  0.20&  \\    
  5&   165.5&     9.77~~~~~& 10.05~~~~~~~~~~&   0.32& triplet\\   
  6&   152.9&     9.69~~~~~& 10.02~~~~~~~~~~&  0.22& \\  
  7&    56.6&     9.68~~~~~& 10.37~~~~~~~~~~&  0.17& \\   
  8&   220.7& $<$10.09~~~~~& 10.13~~~~~~~~~~&   0.21& no spiral members\\   
 10&    64.6&    10.10~~~~~& 10.34~~~~~~~~~~&   0.21& \\   
 14&    73.5&     9.61~~~~~&   9.71~~~~~~~~~~&   0.27&triplet \\   
 15&    91.7&     9.41~~~~~& 10.03~~~~~~~~~~&   0.22& \\    
 16&    52.9& $>$10.42~~~~~& 10.57~~~~~~~~~~&   0.15&  \\   
 18&    54.1&    10.03~~~~~&   9.89~~~~~~~~~~&   0.18& false group \\  
 19&    57.4&     9.31~~~~~&  9.89~~~~~~~~~~&   0.22&  triplet \\   
 20&   195.8& $<$10.06~~~~~&   9.41~~~~~~~~~~&   0.28& \\   
 21&   101.0&    10.36~~~~~& 10.25~~~~~~~~~~&   0.2& 1 \\   
 22&    36.1&     9.13~~~~~&  9.68~~~~~~~~~~&   0.24&triplet \\  
 23&    64.6&    10.03~~~~~& 10.00~~~~~~~~~~&    0.17& \\   
 24&   122.8& $<$ 9.54~~~~~&  9.97~~~~~~~~~~&   0.24& \\   
 25&    85.2&     9.90~~~~~& 10.16~~~~~~~~~~&   0.19& \\ 
 26&   127.3&    10.42~~~~~& 10.23~~~~~~~~~~&   0.16& \\ 
 30&    61.8&    8.62~~~~~& 10.18~~~~~~~~~~&   0.22&   \\  
 31&    54.5&    10.35~~~~~& 10.53~~~~~~~~~~&   0.19& \\ 
 33&   104.6&    10.23~~~~~&  9.48~~~~~~~~~~&   0.30& \\ 
 34&   123.6&    10.70~~~~~&  9.93~~~~~~~~~~&   0.20& \\  
 35&   219.5& $<$10.12~~~~~& 10.01~~~~~~~~~~&    0.22& \\ 
 37&    89.6&     9.19~~~~~& 10.07~~~~~~~~~~&  0.20& \\   
 38&   117.6&     9.69~~~~~& 10.44~~~~~~~~~~&   0.18&triplet \\ 
 40&   89.6&     9.14~~~~~& 10.11~~~~~~~~~~&   0.18& \\   
 42&    53.3&     9.40~~~~~&  9.18~~~~~~~~~~&   0.36& no spiral members\\  
 43&   133.0&    10.11~~~~~& 10.37~~~~~~~~~~&  0.18& \\   
 44&    18.4&     9.23~~~~~&  9.92~~~~~~~~~~&    0.19& \\    
 46&   108.6& $<$ 9.22~~~~~&  9.37~~~~~~~~~~&    0.25& no spiral members\\  
 47&   127.7&     9.93~~~~~& 10.21~~~~~~~~~~&   0.21&   \\ 
 48&    37.7&     8.52~~~~~&  9.34~~~~~~~~~~&    0.22&  \\   
 49&   133.8&    10.54~~~~~& 10.16~~~~~~~~~~&   0.19&   \\    
 51&   103.8& $<$ 9.66~~~~~& 10.17~~~~~~~~~~&   0.17&  \\  
 53&    82.8&    10.04~~~~~& 10.42~~~~~~~~~~&   0.21& triplet \\ 
 54&    19.6&     8.75~~~~~&  9.24~~~~~~~~~~&   0.23& false group \\   
 56&   108.6&     9.36~~~~~& 10.09~~~~~~~~~~&   0.16&     \\   
 57&   122.4&     9.71~~~~~& 10.57~~~~~~~~~~&  0.16&  \\    
 58&    83.2&     9.83~~~~~& 10.40~~~~~~~~~~&   0.19&  \\   
 59&    54.1&     9.49~~~~~&  9.45~~~~~~~~~~&    0.22& no spiral members  \\
 61&    52.1&     9.96~~~~~& 10.04~~~~~~~~~~&    0.20& triplet \\    
 62&    54.9& $<$ 9.06~~~~~&  9.52~~~~~~~~~~&   0.26 & no spiral members \\   
 64&   145.2& $<$10.06~~~~~& 10.29~~~~~~~~~~&   0.19 & triplet \\  
 67&    98.5&    10.03~~~~~& 10.30~~~~~~~~~~&   0.18&  \\    
 68&    32.0&     9.62~~~~~& 10.10~~~~~~~~~~&    0.22&  \\   
 69&  118.4&    10.27~~~~~& 10.18~~~~~~~~~~&   0.17&  \\
 70&   109.9&     9.53~~~~~& 10.16~~~~~~~~~~&  0.20&  triplet \\   
 72&   170.0& $<$ 9.83~~~~~& 10.33~~~~~~~~~~&  0.27 &  \\    
 73&   181.4&    10.84~~~~~& 10.11~~~~~~~~~~&  0.22 & triplet \\    
 74&   161.0& $<$10.08~~~~~&  9.44~~~~~~~~~~&   0.28&no spiral members  \\
 76&   137.0& $<$ 9.51~~~~~& 10.05~~~~~~~~~~&   0.20&  \\ 
\end{tabular}
\end{table*}

\setcounter{table}{1}

\begin{table*}
\caption{HI masses for groups. (cont)}
\label{tab:groupmass}
\begin{tabular}{rrrrcl}
HCG& Distance &
\multicolumn{1}{c}{log[M(HI)$_{\rm obs}$]}&
\multicolumn{1}{c}{log[M(HI)$_{\rm pred}$]$^{a}$}&
s.e.e.&
Remarks\\
& (Mpc)&
\multicolumn{1}{c}{(log[M$_{\odot}$])}&
\multicolumn{1}{c}{(log[M$_{\odot}$])}&
(log[M$_{\odot}$])& \\  
 78&   122.0&    10.40~~~~~& 10.24~~~~~~~~~~&   0.22& triplet \\   
 79&    58.2&     9.30~~~~~&  9.71~~~~~~~~~~&   0.20&  \\    
 80&   124.9&    10.45~~~~~& 10.46~~~~~~~~~~&   0.18&  \\  
 82&   146.0& $<$9.69~~~~~&  10.45~~~~~~~~~~&  0.22&   \\   
 85&   158.6& $<$9.72~~~~~&   9.07~~~~~~~~~~&    0.36 & no spiral members\\    
 87&   116.7&     9.59~~~~~& 10.47~~~~~~~~~~&  0.18 & \\  
 88&    80.7&    10.28~~~~~& 10.55~~~~~~~~~~&  0.17&  \\  
 89&   119.6&    10.45~~~~~& 10.49~~~~~~~~~~&  0.14 &  \\     
 91&    95.7&    10.32~~~~~& 10.56~~~~~~~~~~&  0.17 &  \\    
 92&    86.4&    10.02~~~~~& 10.51~~~~~~~~~~&   0.17&  \\   
 93&    67.4&     9.56~~~~~& 10.36~~~~~~~~~~&  0.24&  \\   
 94&   168.4&    10.22~~~~~& 10.06~~~~~~~~~~&   0.17&  \\ 
 95&   159.8& $>$10.10~~~~~& 10.31~~~~~~~~~~&   0.24& triplet \\    
 96&   117.6& $>$10.20~~~~~& 10.37~~~~~~~~~~&  0.19& \\   
 97&    87.6&     9.10~~~~~&  9.99~~~~~~~~~~&  0.21&  \\  
 98&   107.0&     9.09~~~~~&  9.81~~~~~~~~~~&   0.29 & no spiral members,triplet \\   
 99&   116.7&     9.21~~~~~& 10.11~~~~~~~~~~&  0.26 &  \\  
100&    71.5&     9.74~~~~~& 10.24~~~~~~~~~~&   0.18&  \\    
\end{tabular}
\begin{list}{}{} 
\item[$^{\rm a}$] 
The derivation of the expected HI mass for the 
groups are discussed in \S~\ref{sec:MHISD}.
\end{list} 
\end{table*}

\subsection{Groups mapped with the VLA \label{sec:MHIVLA}}

In Fig. 2 the total observed masses of the 16 compact groups 
imaged with the VLA are compared with the expected one.
The mean deficiency of the groups of the 
VLA subsample is {\it Def$_{\rm HI}$} = 0.16 $\pm$ 0.10 even though
it is biased towards the groups with the strongest HI detections.
The largest deficiency, {\it Def$_{\rm HI}$} = 0.97, is found in
HCG 40, whose detected HI mass is only about 10\% of the
expected mass.
HI excess is inferred for HCG 33, but this group consists mostly of 
ellipticals, and the optical luminosity (thus predicted HI mass)
of the only spiral member H33c is highly uncertain due to a bright
foreground star and large Galactic extinction (b = $-$12.6\ddeg). 
Only lower limits are given for HCG 16, 95 and 96 because
HI absorption strongly affects their spectra.

\begin{figure}
\resizebox{\hsize}{!}{\includegraphics{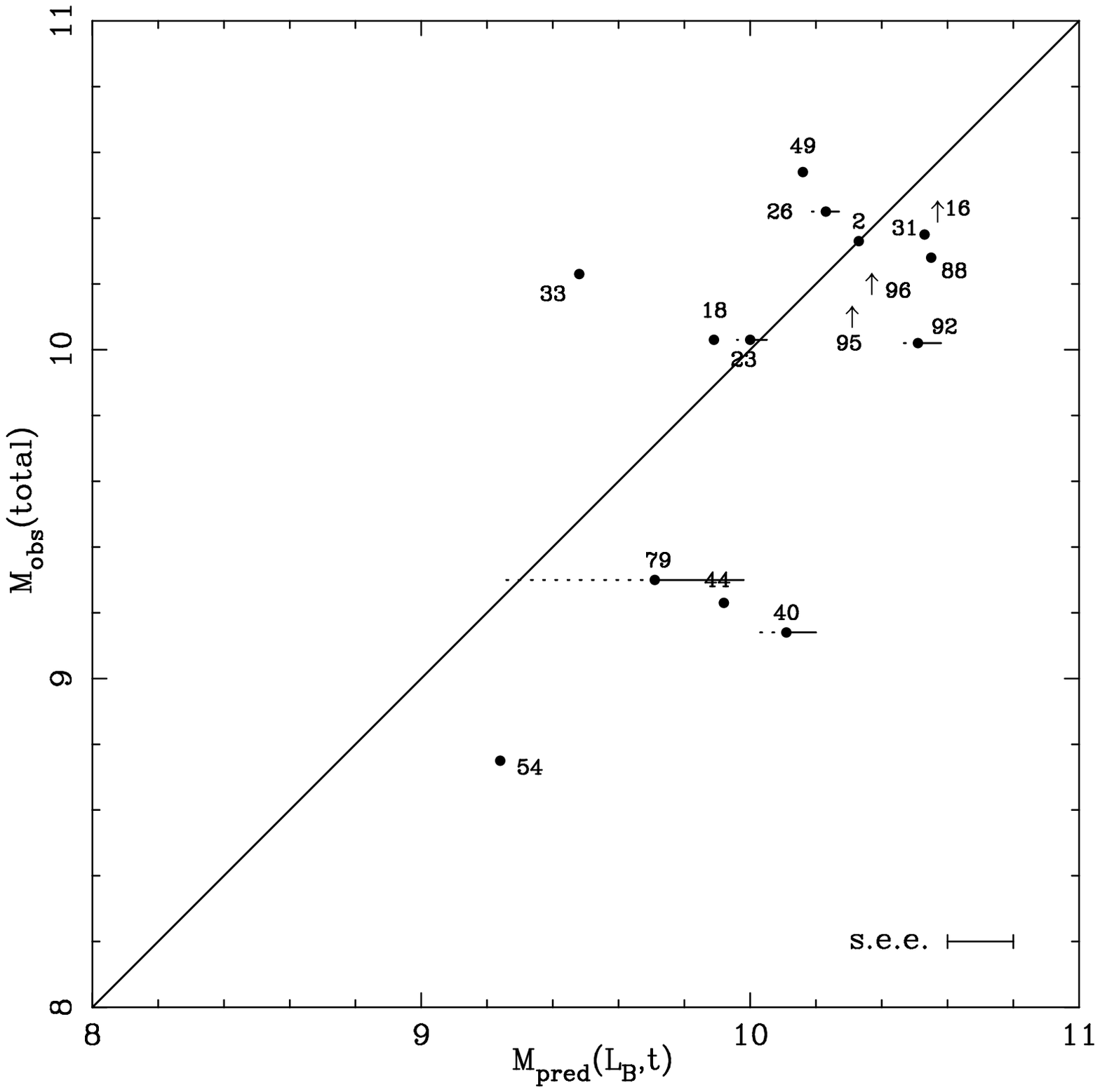}}
\hfill
\caption{Comparison of the total observed HI mass in each
Hickson compact group observed with the VLA with the expected 
one obtained as described in \S~\ref{sec:MHISD}.
Symbols have the same meaning as in Figure~\ref{fig:MHISD}.}
\label{fig:MHIVLA}
\end{figure}

Out of the 70 galaxies (11 ellipticals) included in
the VLA sample of the 16 HCGs, 50 galaxies are detected, 
and the HI emission is resolved for 34 galaxies in 12 groups.
Good upper limits, ranging from 5 $\times$ 10$^{7}$  
to 10$^{8}$ M$_{\odot}$, are obtained for the 9 undetected spiral galaxies. 
We measured the VLA HI flux associated with 
the main body of each galaxy, 
as well as that in gaseous tidal features, bridges and external 
clouds (Table~\ref{tab:galaxymass}). 
Separating HI fluxes for individual galaxies in  HCG 18, 26, 49 and 54 
was not possible because of confusion. 
Fig. 3 shows the observed versus expected HI masses for individual
galaxies, with the
solid line representing a  normal HI content.  
The HI in tidal features has not been included in the observed mass. 
For those cases where a tidal feature could be clearly associated 
to a particular member of a group, a
dotted line connects the HI masses obtained with (open circle) and 
without (filled circle) the tidal feature. Arrows represent upper
or lower limits (due to absorption features in the HI spectra).
The mean HI deficiency of the 43 resolved spiral galaxies
is  {\it Def$_{\rm HI}$} = 0.62 $\pm$ 0.09. 
This is significantly  larger than the mean value for the groups,
due in most cases to
efficient  gas stripping in these galaxies visible in the VLA maps.
The most deficient galaxy is HCG 92b, undetected in HI, with  
{\it Def$_{\rm HI}$} $>$ 2.15, which corresponds to 
0.7\% of the expected mass.

\begin{figure}
\resizebox{\hsize}{!}{\includegraphics{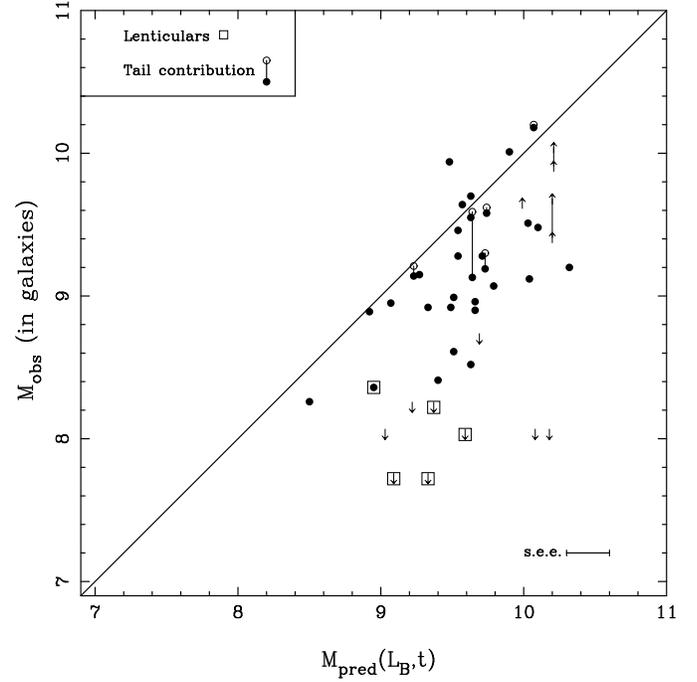}}
\hfill
\caption{Comparison of the measured HI mass in the galaxies of the groups
observed with the VLA with the expected one for their optical
luminosities and morphological types, obtained 
as described in \S~\ref{sec:MHISD}.
Lenticular galaxies are marked with squares. The vertical segments
connect  the masses obtained excluding (filled circle) and
including
(open circle) the tidal feature. Arrows represent upper
or lower limits.}
\label{fig:MHIgal}
\end{figure}

\begin{table}
\caption{HI masses for HCG galaxies.} 
\label{tab:galaxymass}
\begin{tabular}{rrrc}
\multicolumn{1}{c}{Galaxy$^{a}$}&
\multicolumn{1}{c}{log[M(HI)$_{\rm obs}$]}&
log[M(HI)$_{\rm pred}$]$^{b}$&
s.e.e.  \\
&
\multicolumn{1}{c}{(log[M$_{\odot}$])}&
\multicolumn{1}{c}{(log[M$_{\odot}$])}&
(log[M$_{\odot}$]) \\    
  2a&    10.18~~~~~& 10.07~~~~~~~~~~& 0.30\\
  2b&     9.19~~~~~&  9.73~~~~~~~~~~& 0.30\\
  2c&     9.55~~~~~&  9.63~~~~~~~~~~& 0.26\\
 2T&      8.97~~~~~&      ~~~~~~~~~~&     \\
 16a&     9.07~~~~~&  9.79~~~~~~~~~~& 0.38\\
 16b&     8.90~~~~~&  9.66~~~~~~~~~~& 0.38\\
 16c&     9.48~~~~~& 10.10~~~~~~~~~~& 0.30\\
 16d&  $>$9.65~~~~~&  9.99~~~~~~~~~~& 0.30\\
16-3$^{c}$&     9.64~~~~~&  9.57~~~~~~~~~~& 0.38\\
16-6$^{c}$&     8.89~~~~~&  8.92~~~~~~~~~~& 0.36\\
  16T&   10.06~~~~~&      & ~~~~~    \\
 23a&     8.92~~~~~&  9.49~~~~~~~~~~& 0.38  \\ 
 23b&     9.70~~~~~&  9.63~~~~~~~~~~& 0.26\\
 23c&     8.36~~~~~&  8.95~~~~~~~~~~& 0.63\\
 23d&     9.15~~~~~&  9.27~~~~~~~~~~& 0.30\\
 31a&     9.46~~~~~&  9.54~~~~~~~~~~& 0.30\\
 31b&     9.28~~~~~&  9.71~~~~~~~~~~& 0.30\\
 31c&     9.20~~~~~& 10.32~~~~~~~~~~&  0.30\\
 31g&     9.28~~~~~&  9.54~~~~~~~~~~&  0.30\\
 31q&     8.95~~~~~&  9.07~~~~~~~~~~&  0.30\\
 31T&   10.12~~~~~&     &     \\
 33c&     9.94~~~~~&  9.48~~~~~~~~~~& 0.30\\
40b&  $<$8.22~~~~~&  9.37~~~~~~~~~~& 0.36  \\ 
 40c&     8.96~~~~~&  9.66~~~~~~~~~~& 0.29\\
 40d&     8.52~~~~~&  9.63~~~~~~~~~~& 0.38\\
 40e&  $<$8.22~~~~~&  9.22~~~~~~~~~~& 0.26\\
 44a&     8.61~~~~~&  9.51~~~~~~~~~~& 0.38\\
 44c&     8.41~~~~~&  9.40~~~~~~~~~~& 0.26\\
 44d&     8.92~~~~~&  9.33~~~~~~~~~~& 0.30\\
 44e&     8.26~~~~~&  8.50~~~~~~~~~~& 0.26\\
 49T&     8.99~~~~~&       &    \\
 79b&  $<$7.72~~~~~&  9.33~~~~~~~~~~& 0.36\\
 79c&  $<$7.72~~~~~&  9.09~~~~~~~~~~& 0.36\\
 79d&     9.14~~~~~&   9.23~~~~~~~~~~& 0.30\\
 79T&     8.40~~~~~&       &   \\
 88a &     9.12~~~~~&  10.04~~~~~~~~~~& 0.36\\
 88b&      9.51~~~~~  &  10.03~~~~~~~~~~& 0.36\\
 88c&     10.01~~~~~&   9.90~~~~~~~~~~& 0.26\\
 88d&      9.58~~~~~&   9.74~~~~~~~~~~& 0.26\\
 88T&      8.46~~~~~&   &     \\
 92b&   $<$8.03~~~~~&  10.18~~~~~~~~~~& 0.29\\
 92c&   $<$8.03~~~~~&  10.08~~~~~~~~~~& 0.26\\
 92d&   $<$8.03~~~~~&   9.59~~~~~~~~~~& 0.36\\
 92f&   $<$8.03~~~~~&    9.03~~~~~~~~~~& 0.38\\
 92T&       9.98~~~~~&   &    \\
 95c&   $>$9.41~~~~~&   10.20~~~~~~~~~~& 0.30\\
 95d&   $<$8.70~~~~~&    9.69~~~~~~~~~~& 0.26\\
 95T&      9.35~~~~~&     &    \\ 
 96a&   $>$9.91~~~~~&   10.21~~~~~~~~~~& 0.26  \\           
 96c&      9.13~~~~~&    9.64~~~~~~~~~~& 0.26\\
 96d&      8.99~~~~~&    9.51~~~~~~~~~~& 0.30 \\
 96T&      9.73~~~~~&       &    \\
\end{tabular}
\begin{list}{}{}
\item[$^{\rm a}$] A T indicates the HI mass in tidal features external to
the galaxies, as tails and bridges.
\item[$^{\rm b}$] HI mass expected for the group obtained as explained
in Sect. 3.1.
\item[$^{\rm c}$] Notation from De Carvalho et al (1997).
\end{list}
\end{table}

\section{Discussion}

\subsection{Possible systematic effects}

\subsubsection{Early type groups and the role of lenticulars}

Among the 72 groups in our sample, 8 do not contain any spiral member
(HCG 8, 42, 46, 59, 62, 74, 85 and 98)
hence the determination of the HI 
deficiency is highly uncertain. These groups lie mostly in the 
left part of Fig. 1 (large open circles), and when excluded from the 
calculation, the HI deficiency   
rises marginally to {\it Def$_{\rm HI}$} = 0.35  $\pm$ 0.06.
For the rest of the paper these 8 groups will be excluded 
from the 
calculations.

Since  the expected HI
content for S0 galaxies is generally uncertain, we have also calculated 
the decrease in expected HI mass if one assumes them to be devoid of HI
(dashed lines in Fig. 1). In spite of the 
large number of lenticular galaxies in the sample (22\%), the change 
is only significant for a few groups, and the assumption leads to 
a global deficiency of  {\it Def$_{\rm HI}$} = 0.18 $\pm$ 0.06.
On the other hand the early type population excess that exists in HCGs 
(Hickson et al. 1988) could suggest that some of them are
stripped spirals (see e.g. Sulentic 2000). Hence for all groups with 
S0 members we have calculated the increase 
in predicted mass assuming that 
all were originally (conservatively)
Sbs (thick lines in Fig. 1).  
Again this effect would be important just for  a few groups, and 
the mean deficiency would be {\it Def$_{\rm HI}$} = 0.40  $\pm$ 0.06.

\subsubsection{Beam size and HI extent}

We have considered whether the
HI deficiency might be due to a small HI beam diameter (D$_{\rm beam}$).     
If true, the HI deficiency should  be larger for groups 
with larger ratios of the HI extent (D$_{\rm HI}$) with respect to  the 
beam size. Since D$_{\rm HI}$ cannot be derived
from the existing single-dish data, 
we have measured it  for the groups mapped 
at the VLA and compared with their optical extent (D$_{\rm opt}$) 
which is obviously known for the whole sample.
D$_{\rm HI}$ has been determined 
at the lowest common 
HI column density  that can be 
measured (7.7 $\times$ 10$^{19}$ at cm$^{-2}$) 
(Table 1) and the largest  HI extent ($\sim$ 20\arcm\ for HCG 16)
is well within  the VLA primary beam 
of 30\arcm . 
We find that D$_{\rm opt}$ = 0.6 - 1.7 $\times$ D$_{\rm HI}$ and 
all mapped groups  have within 
the optical diameter (D$_{\rm opt}$, given in Hickson 1982 and 
revised here in order to exclude interlopers) at least 85\% of the detected 
HI mass 
which is 
within the intrinsic errors in the determination of M(HI)$_{\rm pred}$. 
Hence, in those cases where the 
single-dish beam is 
equal to or larger than the optical extent of the group 
no significant effect would be expected in the observed HI deficiency.
 This condition is 
met by 70 out of the 72 groups
observed with single dish, and 
the remaining ones, HCG 10 and 21 are not classified as HI deficient.
 This rough calculation suggests  that the beam size did not have a
significant effect on the reported  HI deficiency.

The reasoning above assumes that the total HI extent is given by the 
size measured in the VLA maps. 
Alternatively, 
the HI deficiency could be due to extended but
undetected diffuse HI emission 
either within the 30\arcm\ VLA primary beam, or
extending even 
farther away.
If the ``empty'' area of the VLA beam is filled with part of the
missing HI and the extension of the detected HI fills only a small
percentage of the single-dish beam, then the single-dish flux should be
larger than the VLA flux. 
The premises are 
met for the HI deficient groups but 
no difference is found between the single-dish and VLA fluxes
within the typical sensitivities of the single-dish
measurements, $\sim$ 10$^{8}$ \solmass , 
while $\sim$ 10$^{9}$ - 10$^{10}$  \solmass\
are typically missing in the HI deficient groups (Table 2).
 This result is obtained for only a few groups, although 
including cases of extreme HI deficiency. 
Hence this indicates that the existence of  
significant undetected amounts of diffuse cold gas
does not seem a general or at least the only explanation for the 
HI deficiency.

\subsubsection{Morphological misclassification}

 Estimation of the HI mass expected for a given galaxy depends 
 on its morphology.
Galaxies in compact groups show distortions that can complicate a
reliable determination of their 
morphological type {\it t} (as presented in
Hickson et al. 1989, e.g. {\it t}(Sa) = 3,  {\it t}(Sb) = 5), 
hence a source of
scatter that can mask real trends.
For galaxies earlier than approximately {\it t} = 8 (Scd) errors 
larger than $\Delta${\it t} = 2 would be infrequent, while for later types
errors of the order of  $\Delta${\it t} = 4 
might occur.  The morphology
of 85\% of the
galaxies in the studied groups are earlier than {\it t} = 8,
ranging from E to Sc (Fig. 4), so that large errors 
are not expected. The most probable effect produced by tidal
interactions would be disk stripping, hence favoring 
errors of the order of $\Delta${\it t} = --2. 
We have quantified this effect by comparing the total HI masses
expected for groups where the 
morphological types of its N members are 
{\it t}$_{(i=1,N)}$ and those 
obtained using {\it t}$_{(i=1,N)}$--2 (Fig. 5).
Therefore, in the improbable case that  {\it all } galaxies have been 
misclassified systematically by   $\Delta${\it t} = --2, 
 {\it Def$_{\rm HI}$} would be increased by 0.18.
For a random error of the order of $\pm$ 2  
the spread in the HI deficiency would be
increased by a  similar amount, but
the fraction of deficient groups would be still significant and not  
the result of a wrong determination of morphology.

\begin{figure}
\resizebox{\hsize}{!}{\includegraphics{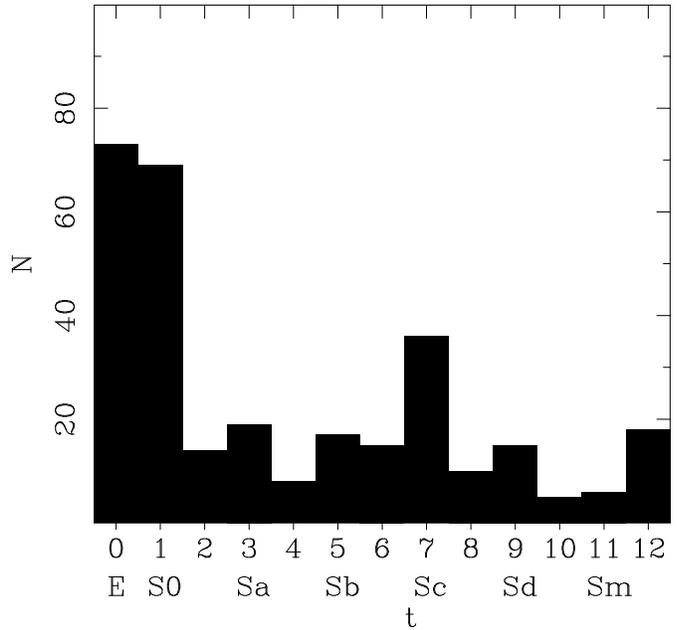}}
\hfill
\caption{A histogram of the morphological types for the 309 galaxies
included in the 72 groups with HI single-dish data.
The numerical assignment of the Hubble types are also given 
under the x-axis. }
\label{fig:histo}
\end{figure}

\begin{figure}
\resizebox{\hsize}{!}{\includegraphics{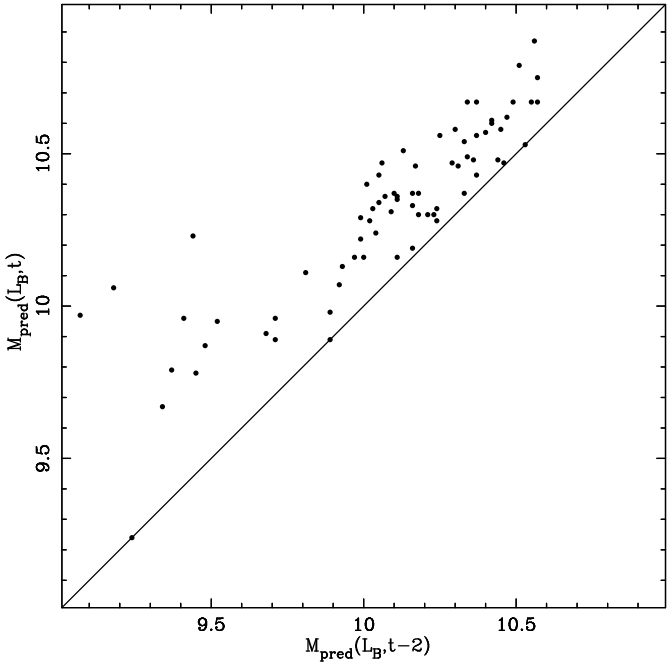}}
\hfill
\caption{Comparison of the total HI masses expected for groups whose 
morphological types are {\it t}$_{(i=1,N)}$ versus those 
obtained using {\it t}$_{(i=1,N)}-2$.  This plot demonstrates the
effect of mis-identifying the Hubble type by $\Delta t = 2$.}
\label{fig:morph_error}
\end{figure}

\subsubsection{Triplets and false groups in the HCG sample}

According to Hickson classification 
(Hickson 1982, Hickson et al 1992), 
a total of 15 triplets exist in the total sample, 
marked with 
smaller symbols in Fig. 1, or 14 in 
our reduced sample of 64 groups
(when groups with no spirals are removed). 
We analyze them separately since 
their physical reality as dynamical entities is not 
clearly established (Thomas \& Couchamn 1992). 
Some of them might be HCGs in formation, 
in a scenario as the one suggested 
by Governato et al. (1996), where groups form and are maintained 
against collapse
by accretion of 
new members. They might also be the result of some merging in 
CGs, but no sign of that is evident from the existing observations.
The HI deficiency of the triple systems is 
not significant, {\it Def$_{\rm HI}$} = 0.19  $\pm$ 0.10, 
which can be understood from the above considerations, as either 
not real systems or unevolved systems.
One of these triplets, HCG 2,  has been mapped by us at the VLA,
and the only signs of interactions found are incipient HI tails arising 
from 2 of its members.
When these systems
are excluded from the sample, the HI deficiency of the 50 groups with
at least 4 members (and at least one spiral) increases marginally to  
{\it Def$_{\rm HI}$} = 0.40  $\pm$ 0.07.

A few cases are found in our sample that might be false
groups, in the sense that instead of being separate 
galaxies they might well be knots of an irregular galaxy,
or at 
most 2 galaxies.
This is the case of HCG 18, which is 
probably 
a single knotty galaxy (see e.g. Williams \& Van Gorkom 1988, 
Plana et al. 2000).  HCG 54 is also  
composed of small optical knots ($\sim$ 0.6 kpc diameter),
embedded in a 
single cloud with a diameter of 12 kpc with  
a long  tail about 20 kpc long. We are currently investigating 
the existence of some neighbor galaxy that might have produced
the tidal tail.
When removed from the sample the deficiency does not change significantly.

\subsection{HI deficiency and group properties}

The most HI deficient groups among those mapped with the VLA  
are HCG 40, 44 and 92, while
the existence of two lenticular galaxies in HCG 79 makes 
its deficiency uncertain. We find 
that the  HI deficiency in them  is  not due to a
few highly anemic objects, but shared by all of their members.
Although this result is obtained based on a small set 
of groups, it suggests that the cause of HI deficiency should be  
associated with a group property.

In this section we explore possible parameters
that can be the origin of the HI deficiency found 
in HCGs. We will exclude from all calculations 
groups with no spirals, triplets and false groups, 
so that the final sample is composed of 48 groups,
with {\it Def$_{\rm HI}$} = 0.40  $\pm$ 0.07.
For the purpose of the statistical calculations 
performed in this section the sample
has been divided between normal HI groups and HI deficient
groups according to the definition given in Sect. 3.1. HI excess
has not been considered since only 3 groups would  
be, in this case, and derived quantities will not have
statistical significance. The rest divides equally 
between normal and deficient groups.

\begin{table}
\caption{Mean physical parameters for the HCGs.}
\label{tab:mean}
\begin{center}
\begin{tabular}{lccc}
Parameter$^{a}$& HI normal & HI deficient & Difference in $\sigma$  \\
\hline 
Fraction of ellipticals &     0.14  &   0.25& 2.2 \\
Fraction of E+S0  &     0.30  &   0.44& 2.0 \\
{\it Comp}/(L$_{\rm B}$ Mpc$^{-2}$)&  390  &   888& 2.5\\
N        &     4.6    &   4.9  & 1.1\\
D$_{\rm G}$ (kpc) &     136.8  &   107.5 & 1.5\\
R$_{\rm 25}$ (kpc)
         &      11.9  &   12.7 & 0.6\\
R$_{\rm p}$ (kpc)&     1.7    &   1.6  & 0.7\\
log $\sigma_v$ (km s$^{-1}$) &     2.14   &   2.31&2.0\\
$\Delta \rho$ 
         &     316    &   557& 1.0\\
n$_{\rm corr}$
         &     1.1  &   2.2 & 1.3\\
{\it Def$_{\rm H_2}$}  
         &    --0.21 &  0.25  &3.2\\
X-ray detected 
         &             4/22 & 10/22 &\\
\\
\end{tabular} 
\begin{list}{}{} 
\item[$^{\rm a}$] The listed parameters are defined in Sect. 4.2.
All of them are mean values of group parameters  calculated on the 
basis
of the final sample of 48 HCGs, except for
{\it Def$_{\rm H_2}$} and for the  X-ray detection rate.
{\it Def$_{\rm H_2}$} is calculated
for the 32 galaxies with both HI and CO data. 
The X-ray detection rate 
is given  for the 44 groups with HI data observed in X-ray by
Ponman et al. (1996).
\end{list} 
\end{center}
\end{table}

\subsubsection{Morphology of the galaxies}

The presence of a 1st-ranked elliptical galaxy, acting 
as a deep potential well,  has been suggested in the
literature as the cause of the HI
deficiency through consumption  or ionization of the
gas  (V\'{\i}lchez \& Iglesias-P{\'a}ramo 1998, Sulentic \& de Mello
Rabaca 1993). From our data the difference in HI deficiency 
between 
groups with a 1st-ranked elliptical  and with a 
1st-ranked spiral,
or even when no elliptical galaxies exist, is not significant.
Therefore a bright elliptical does not appear to play an important
role in the HI deficiency, although 
it cannot be excluded   that this
mechanism is at work in some particular cases. 
Only a marginal (2$\sigma$) trend is found for HI deficient 
groups to be richer 
in early type galaxies (Table 4).

\subsubsection{Group compactness, velocity dispersion, isolation
 and related parameters}

Do the most compact groups show also the largest HI deficiencies?
Since the densest groups  should include the 
most  dynamically evolved, compactness is a parameter
to investigate in this study.
We compared compactness of normal and HI deficient groups 
({\it Comp}, calculated as the  total L$_{\rm B}$ of accordant galaxies 
divided 
by the physical surface area of the group) 
and  found weak (2.5$\sigma$) evidence for the HI deficient groups 
to show higher surface density (Table 4).
No correlation is found between the  number of galaxies in each group 
(N), group diameter (D$_{\rm G}$), galaxy sizes (R$_{\rm 25}$), and 
median projected separation of the galaxies
(R$_{\rm p}$) and the HI deficiency.
HI deficient groups have a marginal (2$\sigma$) trend to have 
larger velocity dispersions ($\sigma_v$)  than normal groups.
 
HCGs were defined with an isolation criterion (Hickson 1982); 
however, new data  have permitted
refinements in the characterization of their environment, showing a 
variety of situations. Measuring on the red POSS plates, Sulentic (1987) 
performed counts of galaxies 
density enhancement  within 0.5$^{\circ}$
of HCGs and within one magnitude of the faintest
galaxy in the group  ($\Delta$$\rho$).
Rood \& Williams (1989) performed similar counts within 
10 radii of the group, and apparent magnitude 
in the range of the group members, correcting for the estimated number
of field galaxies (n$_{\rm corr}$). 
We find no difference in the degree of isolation 
of normal and HI deficient groups
 (Table 4).

\subsubsection{Interstellar medium}

Gravitational  torque during encounters can produce
higher   density conditions due to accretion of HI 
toward the center, inducing 
conversion of atomic into molecular gas  
 (Garc\'{\i}a-Barreto et al. 1994, Menon 1995 and references therein). 
CO data exist for 79 among 202 spirals 
in  44 of the groups with single-dish HI data 
(Verdes-Montenegro et al. 1998,
Leon et al. 1998), and no CO excess  has been found. Instead,
20\% of the groups are in fact CO poor. 
For the 32 galaxies with both CO and  HI data, we have compared
HI and H$_{\rm 2}$ deficiency ({\it Def$_{\rm H_2}$} = 
log[M(H$_{\rm 2}$)$_{\rm pred}$]--log[M(H$_{\rm 2}$)$_{\rm obs}$], 
with M(H$_{\rm 2}$)$_{\rm pred}$
obtained from  Perea et al. 1997). 
We find that galaxies with a larger amount of stripped HI
show a trend to have a depressed CO content   (3$\sigma$, Table 4). 
Quilis et al. (2000) suggested that HI stripping 
would break the balance between 
disruption of
molecular clouds by star formation and condensation of new molecular 
complexes,
so producing H$_{\rm 2}$ deficiency and consequently 
depressed star formation activity.
A general comparison 
between star formation activity (e.g. star formation rate or far-infrared 
emission) and HI content has not been possible due to
the small 
number of galaxies common to 
both samples. 

Sixty-four of the 72 HCGs with HI data have been observed
in X-ray by Ponman et al. (1996), 44 of them belong
to our final sample of 48 (22 normal in HI and 22 deficient), 
and 3 more are only composed of only 
early type galaxies. Among them only 14 have associated 
(diffuse) X-ray emission according to Ponman et al
 (1996).  This leaves us little 
cases for a statistical study of X-ray luminosities. However 
the detection rate is more than double 
for HI deficient groups (Table 4), and we have checked that 
this is not due to larger exposure times
or closer distances. This suggests that 
HI deficient groups are more luminous at X-ray wavelengths. 
Not much information can be obtained
from VLA mapped groups: only 
HCG 33, HCG 16 and HCG 92 have been detected in X-ray and no maps
exist for HCG 33. Recent XMM images of HCG 16 (Turner et al. 2001)
indicate that 
the X-ray emission is not diffuse but 
comes from the nuclei of the galaxies. 
This is consistent with the normal HI content of this group.
In the case of HCG 92 
the anticorrelation between HI and X-ray emission suggests that 
the missing atomic gas was heated in a shock 
after being removed from 
the individual galaxies (see Sect.  4.3).
These evidences are not general enough to be applied to the whole sample.

The largest deficiency among the 72 groups with single-dish measurements
 is found in HCG 30, with 
{\it Def$_{\rm HI}$} = 1.56 (97\%). HCG 30 
is composed of 4 galaxies (3 spirals and 1 lenticular) with 
optical luminosities L$_{\rm B}$ = 7, 5, 3 and 0.8  $\times$ 10$^9$ 
L$_{\odot}$, 
but only 
4 $\times$  10$^8$\solmass\ of atomic gas is detected.
We did not detect CO toward this group,
with an upper limit for each galaxy of 
9  $\times$ 10$^8$\solmass, a factor of 10 lower than expected
for the spirals, the same applying to the FIR luminosity  
(3 $\times$ 10$^9$ L$_{\odot}$ observed versus 
3 $\times$ 10$^{10}$ L$_{\odot}$ expected for each spiral;
Verdes-Montenegro et al. 1998).
Little H$\alpha$ emission is detected in the four galaxies of this group
(V\'{\i}lchez \& Iglesias-P{\'a}ramo 1998),
consistent with the low amount of gas associated with them.
Finally, the density contrast with respect to the environment
is among the lowest. In a radius of 500 kpc four more galaxies with magnitudes
in the range of the group members and velocity differences below
 300\kms\ are listed in NED.

The most extreme cases of HI deficient galaxies ({\it Def$_{\rm HI}$} $>$ 1)
mapped at the VLA
are HCG 92b and c where no HI is detected above 
a 3$\sigma$ value of 0.8\% 
of their expected content.
These galaxies  are strongly deficient in 
molecular gas, with  {\it Def$_{\rm H_2}$} $>$ 1.2 for HCG 92b 
and  {\it Def$_{\rm H_2}$} = 1.0 for HCG 92c (Yun et al. 1997).
The FIR emission of these galaxies is not enhanced with respect to
their optical luminosity,
indicating that disk star formation was not triggered by the interactions.

\subsubsection{Comparison with other dense environments}

We have compared the degree of HI deficiency in HCGs with 
galaxies in Virgo and Coma clusters as well as in the Coma I group.
In Fig. 6 we compare the observed and expected values of HI mass
for these environments. Histograms of the HI deficiency are 
also shown.
The HI data for individual galaxies of HCGs are shown in Fig. 6a 
as filled circles, and the mean value for the groups, 
obtained after normalizing to the
number of members,  as open circles.
In Fig. 6b we plot the data for 313 galaxies in Virgo cluster 
within 10\ddeg 5 of M87 (Huchtmeier 
and Richter 1989).
Data for Coma cluster galaxies are plotted in Fig. 6c (Bravo-Alfaro et al
2000) and for Coma I group in Fig. 6d (Garc\'{\i}a-Barreto et al. 1994). 
In the 3 comparison samples
 the largest deficiencies are reached close to the center
of the cluster/group and with values similar 
 to those found in HCGs, in spite of the different characteristics of 
each environment. 
Deficiencies are larger for Coma I group galaxies 
and  is probably due to a better constraint on the upper limits 
due to its closer distance.
Clusters have larger number of members and velocity
dispersions than groups, and 
ram pressure stripping is the standard explanation there 
for HI deficiency. Coma I cloud is composed of 
 32 galaxies, half of them HI deficient, with  
radial  velocity dispersion similar to compact groups ($\sigma$ = 190
\kms ). Garc\'{\i}a-Barreto et al. (1994) argue that the corresponding 
pressure
of a hypothetical intragroup medium could strip gas from the outer
part of the disks.
Recent numerical simulations such as by Abadi et al. (1999)
have found that ram pressure stripping is not an efficient
mechanism in a group and poor cluster environment, and another mechanism
is likely needed to account for the intragroup medium.

\begin{figure*}
\resizebox{\hsize}{!}{\includegraphics{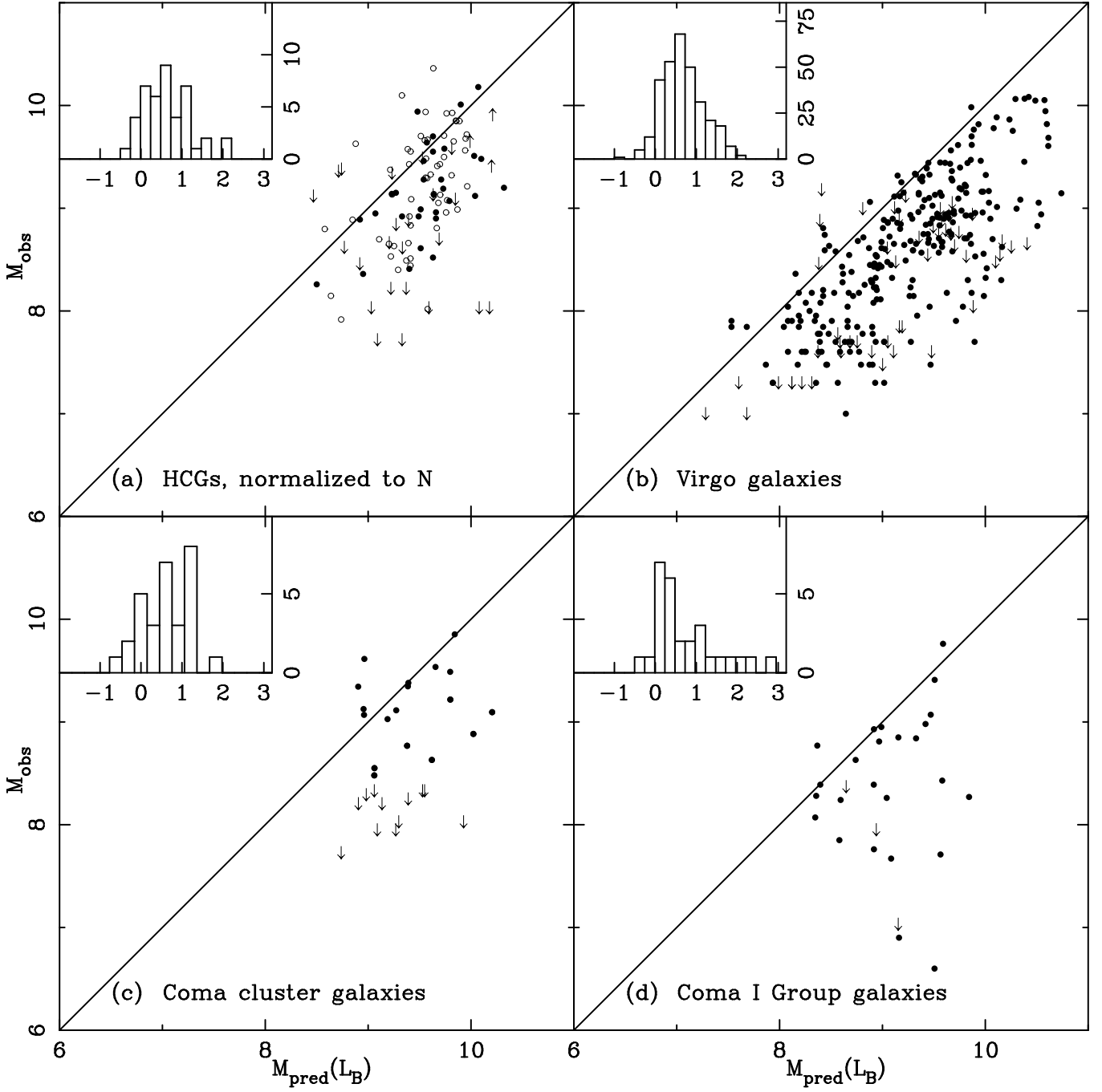}}
\hfill
\caption{Observed versus predicted HI mass for: 
{\bf a} HCGs as a whole (filled circles) where the  total masses  have 
been normalized
to the number of galaxies 
in the group for comparison, and for the
individual galaxies (open circles); {\bf b} Virgo galaxies (Huchtmeier
and Richter 1989); {\bf c} Coma cluster galaxies  (Bravo-Alfaro et al
2000); {\bf d} Coma I group  (Garc\'{\i}a-Barreto et al. 1994).
Histograms of the HI deficiency for each sample are shown 
in the insets of the upper left of each panel.}
\label{fig:HIcompare}
\end{figure*}

The question on how such intragroup medium can be generated
or mimicked in 
the compact group environment 
maybe related 
with the fact that the number of crossings are
much larger than in clusters, where most galaxies
would instead suffer fewer but more violent passages
through  the cluster center. 
A second alternative, where gas is heated by shock
induced by an intruder galaxy  was discussed in section Sect. 4.2.3
in the context of X-ray emission.

\subsection{An evolutionary sequence from the HI distribution}

The groups imaged with the VLA show a 
variety of HI distributions, with 70\% of the spiral galaxies 
perturbed in HI, and their 
morphology and kinematics,
as well as HI content, may be interpreted as different 
stages of group evolution.
There is no one-to-one 
 correspondence between the presence of tidal
features and atomic gas content.
Not all deficient groups show HI tidal tails,
neither are all groups with HI tidal features
deficient in HI, 
e.g. HCG 16, 31, 92 and  96
show numerous tidal tails although HCG 92 is the only
HI deficient.

The normal level of FIR emission found among the HCG galaxies
(Verdes-Montenegro et al. 1997) is not consistent with the usual
scenario of tidally induced gas inflow seen among interacting galaxies
and the contemporaneous, enhanced starburst activity.  Although
the inflow of gas and conversion into stars during the past episodes
of interactions cannot be ruled out, efficient tidal removal of the 
outer HI disks and the inhibition of inflow (see \S~4.2.3) offers a
simpler explanation.
Once the HI is tidally removed from the  individual galaxies,
 the dynamic and energetic compact group environment may be highly 
inducive to dispersal or ionization of the gas and producing 
the observed HI deficiency.  
As discussed by Hibbard et al.
(2000), HI tidal tails are dynamically fragile
and may be easily ionized,
thereby possibly  explaining why some HI deficient groups do not
show tidal tails.  The expected emission measure for
ionized HI tidal tails is of order 1 cm$^{-6}$ pc or less, which
is detectable only by the most sensitive observations. 
This leads us to propose 
a scheme for an evolutionary sequence, shown in Fig. 7 with a sample group 
in each class. 
HCG 95  is not considered here since 
only half of the single-dish HI flux is detected 
in the VLA maps, and found to be located in the galaxies 
 (Huchtmeier et al. 2000).

\begin{figure*}
\resizebox{14cm}{!}{\includegraphics{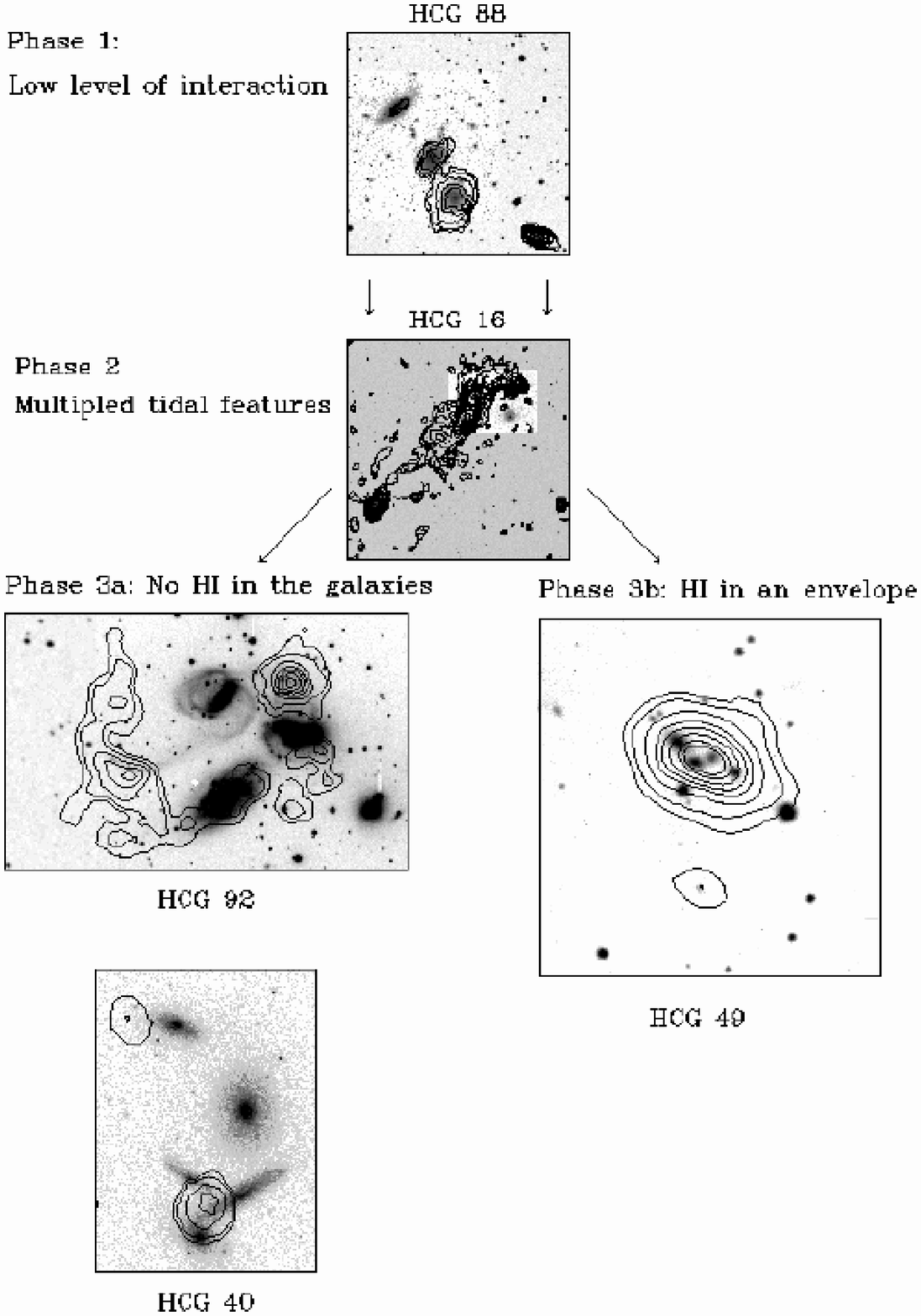}}
\hfill
\caption{A proposed evolutionary scenario for HCGs (see Sect. 4.3). 
The optical image for HCG 88 
is a superposition of a POSS image and a CCD V image obtained with NOT
in La Palma. In the case of HCG 16 an R image (kindly given by 
Iglesias-P{\'a}ramo)
is superposed on a POSS image. 
The R optical image shown for HCG 92 was taken by Sulentic
at the 3.5m telescope of Calar Alto. 
A B image is shown for HCG 49, obtained at the 1.5m telescope of OSN (Granada).
The image of HCG 40 was taken with 
Subaru (Subaru Telescope is operated by the 
National Astronomical Observatory of Japan)
in 1999 (http:www.naoj.org)
and is a combination of J and K$^{\prime}$ frames.
 The shown levels are 
N(HI) = 10, 20, 40, 65, 85, 110, 140, 160, 200, 250, 350, 450, 570
 $\times$ 10$^{19}$ at cm$^{-2}$.}
\label{fig:evolution}
\end{figure*}

In Phase 1 the HI distribution and kinematics are
relatively unperturbed, and more  than $\sim$ 
90\% of the HI mass is located in the disks of the
galaxies with the remaining gas found in incipient tidal
tails. Examples of this are HCG 88 (top of the tree in Fig. 7)
together with HCG 23 and 33.
HCG 88 has a
 high density of galaxies
(160 galaxies/Mpc$^2$) and high degree of
isolation (only two galaxies, 4 magnitudes fainter than galaxy d, are found in 
a 700pc-side square 
centered on the group; De Carvalho et al. 1997), and  
is especially interesting since this
quartet of spirals, aligned in projection, has been  considered 
a good candidate for a projected filament (Hernquist et al. 1995). 
However the properties of this group, in particular velocity dispersion 
($\sigma_v$ = 27 km s$^{-1}$) and crossing time 
(H$_{\rm 0}~t_{\rm c}$ = 8.70) , are quite different from those predicted 
by the simulations performed by Hernquist et al. (1995; see
their Table 1). 
Therefore we think that HCG\, 88 is a good example of a physically dense group 
at a very early stage of interaction.

In Phase 2, groups still retain a 
significant amount of HI in the disks of the spirals but 
30 to  60\% of the
total HI mass forms   tidal features. The HI morphology in these groups can be 
understood as an extrapolation of interactions in pairs of galaxies,
where tails and bridges are often seen.
 Examples are HCG 96, 16 and 31, 
ordered from lower to higher percentage of gas in tidal features. In the
figure we show HCG 16 as an example where the intragroup atomic gas composes 
 a grid of  bridges and tails.

In Phase 3a most if not all of the HI
have been stripped from the disks of the galaxies
and are either found in  
tails, produced in multiple tidal interactions, or not 
even detected. Examples of this are HCG 40, 44 and 92. 
HCG 92 is a specially striking case since all the observed HI  
 (35\% of the expected content) has been stripped out from the galaxies 
 and formed several 
intragalactic clouds and tidal tails (Williams et al. 2001).
Abundant 
heated and/or ionized gas is observed: X-ray emission 
from shocked gas (Pietsch et al. 1997)
anticorrelates
with the HI, 
and intense bursts of star formation external to the galaxies 
have been reported by several
authors (e.g. Moles et al. 1997, Plana et al. 1999, Xu et al. 1999, 
Gallagher et al. 2000). 
The most extreme case is HCG 40, where HI is only detected
in a fraction of the disks of two of its spiral members, 
and no HI external to the galaxies is detected, so 
that 10$^{10}$\solmass\ (90\% of the expected HI) are missing.
We think that the HI was stripped and then suffered more easily 
a phase change, although this requires further 
confirmation since no X-ray measurement exists for this group.

The least common Phase (named 3b in Fig. 7) involves groups where the 
 HI gas forms a large cloud containing all galaxies, and with  
a single velocity gradient.  The total HI spectrum
has a characteristic single peaked shape.
This appears to be a rare situation, since among the 72 groups
observed with single dish (excluding HCG 18 and HCG 54 as probable
false groups, see Sect. 4.1.4) only HCG 26 (Williams \& Van Gorkom 1995) and 
49 show
this kind of spectrum.  HCG 26 is dominated by a large edge-on spiral
whose HI is difficult to differentiate from the whole cloud.  HCG 49
is the best defined and most impressive example of this phase, shown
in the right branch of Fig. 7.
The HI cloud has an elliptical shape with a size of 95 $\times$ 65 kpc, 
and a velocity
gradient of $\sim$ 250~km~s$^{-1}$, containing 4 well differentiated
galaxies with a velocity dispersion as low as 34~km~s$^{-1}$ and a
total optical diameter of 35 kpc. 
The morphology and kinematics of this cloud (Fig. 8a) is not likely 
 the result of smoothed emission from multiple tidal tails, like
those found  in HCG 31 (Fig. 8b).
Smoothing the HCG 31 datacube 
to the same physical resolution as in HCG 49 (Fig. 8c) gives a similar
overall morphology  
but different kinematics, since no continuous velocity gradient
is obtained for HCG 31 with a lower resolution.
The galaxies of this group are similar in luminosity and type to M33, 
for which 
extended HI envelopes are not unusual (Huchtmeier 1973). 
More extended atomic gas would be more easily 
detached from the individual 
members  to form  a common HI cloud
(A. Bosma, priv. comm.). 
 Hence evolution from Phase 2 to 3a or 3b 
would depend on the size of the galaxies. Small optical disks
with respect to the HI extent may favor the formation
of a single cloud.
 We are currently performing optical imaging
and spectroscopy of
this group in order to better characterize the dynamnics of the group.

\begin{figure*}
\resizebox{\hsize}{!}{\includegraphics{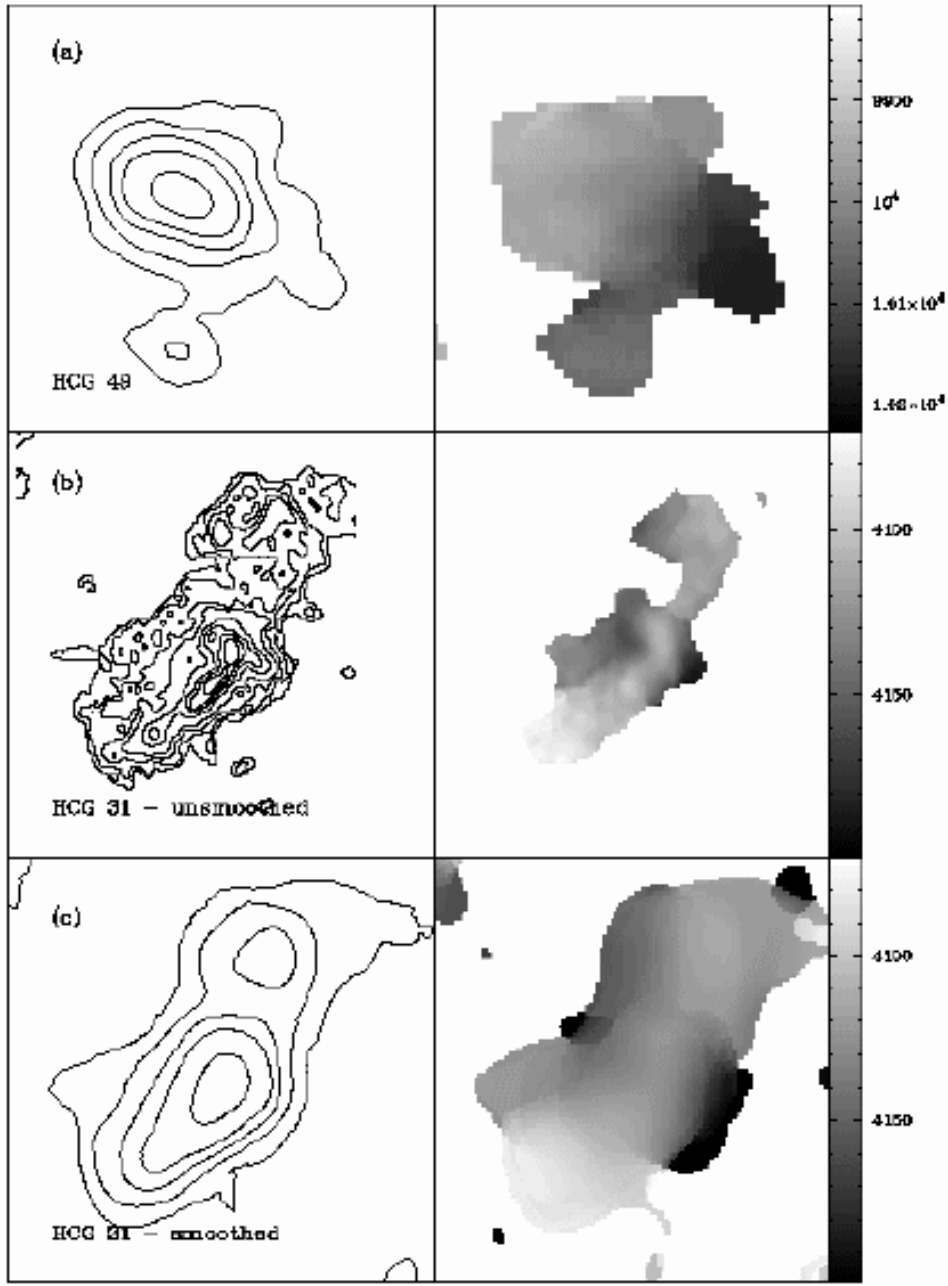}}
\vspace*{-2cm}
\hfill
\caption{A test of resolution effects.
{\bf a} HI integrated emission (left) and velocity field (right) of HCG 49.
{\bf b} The same as in  {\bf a} for HCG 31. {\bf c} The same as in {\bf b} 
smoothed to
a beam with the same physical size as in HCG 49 (63\arcsec\ $\times$
52\arcsec ). The shown levels are 
N(HI) = 4, 18, 40, 72, 143, 206, 252, 295 $\times$ 10$^{19}$ at cm$^{-2}$.}
\label{fig:resolution}
\end{figure*}

The case of HCG 79 (Seyfert's sextet) is considered here 
separately due to the uncertainty in its classification. 
This group  
contains a giant elliptical with a prominent dust lane
as well as FIR and radiocontinuum emission, 
two lenticulars and a single spiral
galaxy, from which an HI and 2 optical tails are emerging.
However the signs of  
dynamical evolution are stronger in the optical, 
with an extended luminous halo
(Sulentic \& Lorre 1983, Nishiura et al. 2000). The detected HI 
tail might have been produced via interaction of the spiral member
with this halo.

\section{Conclusions}

A combined analysis of single-dish data for 72 HCGs
and high resolution mapping for 16 shows:
\begin{itemize}
\item 
The full sample of 72 groups
with HI single-dish data has a mean HI deficiency of
 {\it Def$_{\rm HI}$} = 0.36 $\pm$ 0.06. When a more homogeneus
sample is extracted, excluding 
 triplets,
false groups, and groups with no spiral members, the remaining 48 groups 
show a larger mean HI deficiency of 
0.40 $\pm$ 0.07.
These values  depend only weakly on expected
errors in the classification of the morphological types.
\item
The galaxies mapped with the VLA have a larger 
degree of deficiency, {\it Def$_{\rm HI}$} = 0.62 $\pm$ 0.09 
in agreement with the efficient gas stripping that takes place,
evidenced by HI tails and bridges.  
\item We have investigated whether the deficiency could be 
due to HI missed from the beam, either due to a small 
beamsize relative to the group extent, or to the presence 
of diffuse undetected emission.
If  the HI extent observed in the VLA maps is 
reasonably representative
of the whole sample, then the beam size did not have a
relevant effect on the reported  HI deficiency.   
We discard also the presence of significant 
amounts of extended cold gas that escaped the observations, 
from comparison between VLA and single-dish fluxes.
\item
A large set of group parameters is explored as the origin of the 
HI deficiency, 
including morphological types, size of the groups and galaxies,
compactness, velocity dispersion, isolation and ISM.
We  find a weak (2 - 2.5 $\sigma$) trend for the HI deficient groups to be 
early type rich, more compact, and show higher velocity dispersions.
A slightly stronger correlation exists (3 $\sigma$ level)
with the CO content: HI deficient groups have also depressed CO emission.
The study of a larger and complete sample (as the one defined by 
Pradoni et al 1994, see also  Coziol et al. 2000)
would be desirable 
to confirm these trends.
\item The most extreme case of deficient group (HCG 30; {\it Def$_{\rm HI}$} = 
1.56 or  97\% of the expected HI missing) as well
as deficient galaxies (HCG 92b and d; {\it Def$_{\rm HI}$} = 2.1 -- 99\%)
show also a depressed content of molecular gas and no FIR enhancement.
Star formation seems therefore inhibited by intense HI stripping
 from the galactic disks that breaks the ISM equilibrium.
\item The HI deficiency level is similar to the one found in 
the central galaxies of Coma and Virgo clusters, as well as Coma I
group, although there the deficiencies seems to be related to
the existence of a hot intragroup medium. The existence 
of a similar mechanism in groups is not clear.
Among the 44 groups of our sample observed in X-ray, the detection 
rate is higher for the HI deficient groups, and this is not due to 
closer distances or longer exposure times.
However the distribution of the X-ray emission is not clear.
HCG 16 shows only nuclear X-ray emission while  HCG 92 
shows heated gas
cooling at X-ray wavelengths, with  the hot 
gas anticorrelated with the neutral atomic gas.
\item
The  mapped groups show a
variety of HI distributions, with 70\% of the spiral galaxies
perturbed in HI. These results are combined in 
 a possible evolutionary scenario.
Least evolved groups are found with a low level of interaction.
Later,  multiple tidal tails form, a stage that can 
be understood as an extrapolation of  pair interactions. 
Once the gas is removed from the galaxies it can be more 
easily heated or destroyed, thereby producing HI deficient groups.
In a few cases,  the HI evolves to form a single cloud containing 
the whole group, 
which might be related 
to 
the smaller size of the galaxies.
\end{itemize}

The general process producing HI  deficiency in a large number of HCGs
 is yet to be clarified.
Our results suggest that gas heating maybe playing a role, 
but  more sensitive X-ray fluxes and maps
 for a larger set of groups 
with available HI data would substantially help to improve
our understanding.

\begin{acknowledgements}
LVM acknowledges interesting discussions with J. Sulentic,
P. T. P. Ho,
R. Sancisi, E. Athanassoula and A. Bosma. LV--M, AO and JP are
partially supported by DGI (Spain)
Grant AYA2000-1564 and Junta de Andaluc\'{\i}a (Spain).
\end{acknowledgements}

\end{document}